\documentclass{article}

\usepackage{arxiv}

\usepackage[utf8]{inputenc} % allow utf-8 input
\usepackage[T1]{fontenc}    % use 8-bit T1 fonts
\usepackage{hyperref}       % hyperlinks
\usepackage{url}            % simple URL typesetting
\usepackage{booktabs}       % professional-quality tables
\usepackage{amsfonts}       % blackboard math symbols
\usepackage{amsmath}        % helpful equation environments
\usepackage{xfrac}          % compact symbols for 1/2, etc.
\usepackage{amsthm}         % for theorems, corollaries, and lemmas
\usepackage{microtype}      % microtypography
\usepackage{siunitx}        % standard international units formatting
\usepackage{color}
\usepackage{graphicx}
\usepackage{natbib}
\usepackage[sectionbib]{bibunits}       % separate bib for appendix
\defaultbibliographystyle{abbrvnat}
\defaultbibliography{references}
\usepackage{doi}

\usepackage{float} %Added by Hani

\newtheorem{theorem}{Theorem}
\newtheorem{proposition}{Proposition}
\newtheorem{corollary}{Corollary}[theorem]

\title{Assumption-Lean Differential Variance Inference for Heterogeneous Treatment Effect Detection}

\date{}

\author{Philippe A. Boileau \\
    McGill University, and \\
    Research Institute of the McGill \\
    University Health Centre\\
    \texttt{philippe.boileau@mcgill.ca} \\
    \And
    Hani Zaki \\
    Universit\'{e} de Montr\'{e}al \\
    \texttt{hani.zaki@umontreal.ca} \\
    \And
    Gabriele Lileikyte \\
    Lund University \\
    \texttt{gabriele.lileikyte@med.lu.se} \\
    \And
    Niklas Nielsen \\
    Lund University \\
    \texttt{niklas.nielsen@med.lu.se} \\\
    \And
    Patrick R. Lawler \\
    McGill University, \\
    Research Institute of the McGill \\
    University Health Centre, and \\
    University of Toronto \\
    \texttt{patrick.lawler@mcgill.ca} \\
    \And
    Mireille E. Schnitzer \\
    Universit\'{e} de Montr\'{e}al, and \\
    McGill University \\
    \texttt{mireille.schnitzer@umontreal.ca} \\
}

\begin{document}
\maketitle

\begin{abstract}
  The conditional average treatment effect (CATE) is frequently estimated in
  clinical studies to refute a homogeneous treatment effect hypothesis. Under
  this regime, all patients making up the population experience
  identical benefit from a given treatment relative to a comparator.
  Uncovering heterogeneous treatment effects through inference about the CATE,
  however, requires that covariates truly modifying the treatment effect be
  reliably collected at baseline. CATE-based techniques will necessarily fail to
  detect violations when effect modifiers are omitted from the data due to, for
  example, resource constraints. Severe measurement error has a similar impact.
  Clinical decision makers can be misled as a result. To address these
  limitations, we prove that a practical homogeneous treatment effect hypothesis
  can be gauged through inference about contrasts of the potential outcomes’
  variances even when effect modifiers are missing from the data. We derive causal
  machine learning estimators of these contrasts and study their asymptotic
  properties. We establish that these estimators are double robust and
  asymptotically linear under mild conditions, permitting formal hypothesis testing
  about treatment effect heterogeneity. Numerical experiments demonstrate
  that these estimators’ asymptotic guarantees are approximately achieved in
  finite-sample randomized and observational study data alike. These inference
  procedures are then used to detect heterogeneous treatment effects in the
  re-analysis of randomized controlled trials investigating targeted temperature
  management in cardiac arrest patients.
\end{abstract}

% keywords can be removed

\begin{bibunit}

\section{Introduction}\label{sec:intro}

The canonical analysis of a continuous outcome in a parallel group design takes the
average treatment effect (ATE) as the target of inference. The ATE is defined as
the difference of the potential outcomes' means, where potential outcomes are
the outcomes that would be observed had study units been assigned to the
treatment or control groups \citep{neyman1923,rubinEstimatingCausalEffects1974}.
While the difference of the potential outcome distributions' first moment is
widely accepted as a useful summary of the individual treatment effect
distribution, it only provides a complete characterization of this distribution
under a shifted version of the sharp homogeneous treatment effect hypothesis
\citep{fisher1935design}. This hypothesis states that the individual treatment
effects are constant in the population under study.

Practitioners in numerous fields have readily acknowledged that the ATE provides
a generally insufficient comparison of competing treatments' effects. Decisions
informed solely by inference about the ATE can overlook the beneficial ---
or detrimental --- effects of a novel treatment in certain subgroups. In medicine,
for example, myriad clinical trials have failed to show that an
investigational therapy meaningfully improves patient outcomes relative to
the standard of care on average, yet have provided evidence of a differential
effect in some patient phenotypes \citep[see, for instance, the following oncology
trials:][]{buttsTecemotideLBLP25Placebo2014, angRandomizedPhaseIII2014,
  zhuRamucirumabPlaceboSecondline2015}.

Numerous methods for quantifying treatment effect heterogeneity have therefore
been developed to supplement inference about the ATE. Chief among them are
inference procedures based on the conditional average treatment effect (CATE)
\citep[see][a non-exhaustive list of
examples]{luedtkeSuperLearningOptimalDynamic2016,
  chenGeneralStatisticalFramework2017,
  wagerEstimationInferenceHeterogeneous2018,
  chernozhukovGenericMachineLearning2018, levyFundamentalMeasureTreatment2021,
  zhao2021, bahamyirouDoublyRobustAdaptive2022, kennedyOptimalDoublyRobust2023}.
The CATE, defined as the expected difference of potential outcomes conditional on
baseline covariates, can be used to identify subpopulations that 
experience differential average benefit as a function of their
covariates. As a result, the CATE can only uncover effect heterogeneity if the
treatment effect modifiers are included among the pre-treatment covariates in
its conditioning set. CATE-based procedures are therefore unable to detect
treatment effect heterogeneity when effect modifiers are inadvertently
excluded from the data owing to resource constraints --- or are included but
severely mismeasured due to measurement instrument limitations. Alternatively,
effect modifiers may also be overlooked during data collection when the mechanistic
understanding of the intervention is incomplete. Health care professionals
relying on CATE inference to inform patient care decisions or develop treatment
guidelines may therefore be misled, confusing an absence of effect modifiers for
evidence of treatment effect homogeneity.

Including a large number of potential effect modifiers in the CATE's
conditioning set may not be a reasonable strategy for uncovering drivers of
treatment effect heterogeneity either. \citet{kennedyOptimalDoublyRobust2023} proved that
high-dimensional inference about the CATE is unreliable unless unverifiable
simplifying assumptions are made about the DGP, like sparsity. Recent work by
\citet{boileauFlexibleApproachPredictive2023, boileauGuidanceIndividualizedTreatment2025}
provided additional empirical evidence that inference procedures for the CATE produced
elevated false positive and negative rates when attempting to identify treatment effect
modifiers from among many pre-treatment covariates.

Inference procedures for treatment effect modifier variable importance
parameters (TEM-VIPs) offer an alternative avenue for detecting treatment effect
heterogeneity \citep{williamsonNonparametricVariableImportance2021,
  boileauFlexibleApproachPredictive2023,
  williamsonGeneralFrameworkInference2023, hinesVariableImportanceMeasures2023,
  boileauNonparametricFrameworkTreatment2025,
  ziersenVariableImportanceMeasures2025}. Given a set of baseline
covariates, these approaches evaluate variables' individual capacity for
treatment effect modification using variable importance parameters. But these TEM-VIP
procedures, like approaches targeting the CATE, can only uncover effect
heterogeneity when the effect modifiers are included in the data.

Further, inference about the CATE and TEM-VIPs will necessarily disregard
residual heterogeneity --- that is, when the expected
differences of the potential outcomes are identical, up to a constant,
conditional on available baseline covariates, but their variances are different.
This information is particularly relevant to medical decision making, as an
intervention may be preferred over another that produces identical results
on average but generates more variable outcomes.

All together, these limitations make the discovery and quantification of
heterogeneous treatment effects challenging in clinical contexts where potential
effect modifiers are missing. A primary example of this is in critical care
medicine. The reliable measurement of baseline covariates for studies in this
medical setting is difficult owing to patient acuity and complexity, possibly
limiting practitioners' ability to rely on data-driven, precision medicine
evidence to guide patient care.

These challenges are exemplified by this work's motivating application. Targeted
temperature management, an intervention that induces mild, controlled
hypothermia, was historically indicated for comatose adults following cardiac
arrest. Pre-clinical and observational studies had previously suggested that
this intervention could improve these patients' neurological outcomes
\citep{hirschPart11Post2025}. At the time, however, there was clinical
uncertainty regarding the target temperature that optimized patient outcomes.
The Target Temperature Management at 33°C versus 36°C after Cardiac Arrest (TTM;
NCT01020916) and Hypothermia versus Normothermia after Out-of-hospital Cardiac
Arrest (TTM2; NCT02908308) randomized controlled trials (RCTs) investigated
whether various target temperatures improved the six-month overall survival of
unconscious patients resuscitated from out-of-hospital cardiac arrest
\citep{nielsenTargetedTemperatureManagement2013,
  dankiewiczHypothermiaNormothermiaOutofHospital2021}. Neither trial reported a
statistically significant difference between their respective arms' outcomes on
average. As a result, contemporary guidelines recommend maintaining patients'
temperatures between \SI{32}{\celsius} and \SI{37.5}{\celsius}
\citep{hirschPart11Post2025}. Nevertheless, clinical uncertainty
remains. \citet{markPersonalizingTemperatureTargets2025} recently hypothesized
that certain clinical traits --- such as severity of neurological injury on
admission and duration of ischemia --- may modify the effect of target
temperature management. Investigating this hypothesis is challenging, however,
due to difficulties ascertaining potentially key baseline characteristics. For
example, the time to initiation of advanced life saving support, a covariate
closely related to ischemia duration, was not recorded directly in the TTM and
TTM2 trials, but derived from emergency medical services and hospital notes.
Indeed, forthcoming work has attempted to uncover this possible heterogeneity
using several CATE and TEM-VIP inference strategies to no avail
\citep{lileikyte2026}.

To address the limitations of conventional approaches, we develop
assumption-lean inference procedures 
\citep{vansteelandtAssumptionleanInferenceGeneralised2022} for detecting 
heterogeneous treatment effects that do not rely on access to effect modifiers. 
This procedure depends on contrasts of the potential outcomes' variances,
which we refer to as \textit{differential variance} estimands, and can be
used to evaluate a relaxed version of the sharp homogeneous treatment effect 
hypothesis. We then derive one-step 
\citep{pfanzaglCONTRIBUTIONSGENERALASYMPTOTIC1985,
  bickelEfficientAdaptiveEstimation1998} and targeted maximum likelihood
estimators (TMLE) \citep{vanderlaanTargetedLearningCausal2011,
  vanderlaanTargetedLearningData2018} of these causal contrasts based on their
efficient influence functions (EIFs). We study these estimators' asymptotic
behavior, showing that they are double robust and providing conditions under
which they are asymptotically linear in randomized and observational studies.

The remainder of the article is organized as follows:
Section~\ref{sec:problem-formulation} formalizes the general causal and
statistical inference problems, linking the two through a causal identifiability
result. Section~\ref{sec:inference} introduces EIF-based estimators of the
potential outcomes' variances and details their asymptotic properties. Inference
procedures for differential variance estimands are also described.
Section~\ref{sec:simulations} showcases simulation studies that empirically
validate the asymptotic results of the earlier sections.
Section~\ref{sec:application} applies the proposed methodology in the
motivating application. Finally, Section~\ref{sec:discussion} discusses the
proposed methodology and outlines possible extensions.

\section{Problem Formulation}\label{sec:problem-formulation}

\subsection{Differential Variance Estimands}\label{subsec:full-data-dgp}

Let the full-data DGP consist of $n$ independent and identically distributed
(i.i.d.) random vectors
$X_i = (W_i, A_i, Y^{(0)}_i, Y^{(1)}_i) \sim P_{X,0} \in \mathcal{M}_X$ for
$i = 1, \ldots, n$. We drop the indices where convenient throughout the
remainder of the text. Here, $W$ is a random vector of pre-treatment covariates
that contains $U$, the random vector of \textit{measured} treatment--outcome
confounders, and $V$, the random vector of \textit{measured} 
effect modifiers. The elements of $U$ and $V$ are not mutually
exclusive and can be empty, implying that true confounders or effect modifiers,
if any, are possibly unmeasured. $A$ is a random binary variable with
support $\mathcal{A} = \{0, 1\}$ representing treatment assignment; observations
are said to be assigned to the treatment group when $A=1$ and to the control
group otherwise. $Y^{(0)}$ and $Y^{(1)}$ are potential outcomes, continuous
random variables representing the outcome under assignment to the control and
treatment groups, respectively. Finally, $P_{X,0}$ is
the true, but unknown, full-data DGP, an element of the nonparametric
statistical model $\mathcal{M}_X$ containing all possible full-data DGPs.

While realizations of $X$ are generally unobservable owing to the definition of
potential outcomes, the full-data DGP permits the definition of parameters that
admit causal interpretations. In particular, we wish to construct a causal estimand
that can be used to detect treatment effect heterogeneity or, equivalently,
uncover departures from a treatment effect homogeneity hypothesis without
requiring access to $V$.

Consider the shifted sharp homogeneous treatment effect hypothesis, where
$Y^{(1)} = Y^{(0)} + \gamma$ for
$\gamma = \mathbb{E}_{P_{X,0}}[Y^{(1)}-Y^{(0)}] \in \mathbb{R}$, the ATE. Then a
natural target of inference is the difference of the potential outcome
standard deviations, which we refer to as the
\textit{absolute differential variance} estimand:
\begin{equation*}
    \Psi_F(P_{X,0}) = \sigma_F(P_{X,0}; 1) - \sigma_F(P_{X,0}; 0) \;,
\end{equation*}
where
\begin{equation*}
  \sigma^2_F(P_{X,0}; a) = \mu^2_F(P_{X,0}; a) - \mu_F(P_{X,0}; a)^2 \;.
\end{equation*}
Here, $\mu_F^{d}(P_{X,0}; a) = \mathbb{E}_{P_{X,0}}[(Y^{(a)})^d]$ for
$d \in \{1, 2\}$, $\sigma_F(P_{X,0}; a) = \sqrt{\sigma^2_F(P_{X,0}; a)}$, and the
subscript $F$ is used to denote full-data DGP parameters. To simplify notation,
we omit the superscript $1$ when $d=1$ throughout the article. This causal
estimand provides an interpretable, absolute contrast of the potential
outcomes' variability. Positive values, which are on the same scale as the potential
outcomes, indicate that the potential outcomes under treatment have greater
variance than the potential outcomes under control, and negative values the
opposite. Further, by the properties of the variance, $\Psi_{F}(P_{X,0}) \neq 0$
implies that $Y^{(1)} \neq Y^{(0)} + \gamma$ so that the shifted sharp homogeneous
treatment effect hypothesis cannot hold. Importantly, $\Psi_{F}(P_{X,0})$
is a marginal parameter that does not need to condition on $V$ to capture
heterogeneity. In contrast, the
CATE will necessarily fail to detect violations of the homogeneous treatment
effect hypothesis by conditioning on $W$ if $V = \emptyset$ --- that is,
when the effect modifiers driving the heterogeneous effect are not included in
the data such that $\mathbb{E}_{P_{X,0}}[Y^{(1)}-Y^{(0)}|W] = \gamma$ almost surely
(a.s.) --- even though $Y^{(1)} \neq Y^{(0)} + \gamma$ .

This sharp homogeneous treatment effect hypothesis is restrictive,
however, suggesting that its violations are not always informative of
meaningful treatment effect heterogeneity. To see this, let
$Y^{(a)} = f^{(a)}(W) + \epsilon^{(a)}$ where
$f^{(a)}(W) = \mathbb{E}_{P_{X,0}}[Y^{(a)}|W]$ for $a \in \mathcal{A}$ and
$\epsilon^{(a)}$ is a mean-zero, possibly heteroscedastic residual noise
term capturing the effect of unmeasured prognostic variables and effect
modifiers, if any. Then, under this notion of homogeneity,
$f^{(1)}(W) = f^{(0)}(W) + \gamma$ a.s. and
$\epsilon^{(1)}=\epsilon^{(0)}$. The shifted sharp homogeneous treatment
effect hypothesis therefore imposes the constraint that the
potential outcomes' residual noise terms be equal. Clinical studies
inadvertently comparing two equivalent interventions such that
$f^{(1)}(W) = f^{(0)}(W)$ a.s. but $\epsilon^{(1)}$ and $\epsilon^{(0)}$
are i.i.d. do not satisfy this restrictive definition since
$\epsilon^{(1)} \neq \epsilon^{(0)}$.

We instead consider a more pragmatic concept of homogeneity,
which we call the \textit{second-moment} homogeneous treatment effect
hypothesis. We define a treatment effect as second-moment homogeneous if
$f^{(1)}(W) = f^{(0)}(W) + \gamma$ a.s. and
$\mathbb{E}_{P_{X,0}}[\mathbb{V}_{P_{X,0}}[\epsilon^{(1)}|W]] = \mathbb{E}_{P_{X,0}}[\mathbb{V}_{P_{X,0}}[\epsilon^{(0)}|W]]$.
This version of homogeneity loosens the second constraint of the sharp
hypothesis, permitting the potential outcomes' residual noise terms to be
different so long as the means of their conditional variances are equal.
When the noise terms are homoscedastic, this condition simplifies to 
$\mathbb{V}_{P_{X,0}}[\epsilon^{(1)}] = \mathbb{V}_{P_{X,0}}[\epsilon^{(0)}]$.

This relaxation is intuitive: potential outcomes with different mean residual
noise term variances cannot induce a constant individual treatment effect,
reflecting the standard definition of effect heterogeneity. Indeed,
differences in the average variances of $\epsilon^{(1)}$ and $\epsilon^{(0)}$,
when present, are attributable to systematic variation created by missing or
mismeasured effect modifiers. The contribution of unmeasured prognostic variables
to the mean conditional noise term variances is otherwise identical by the definition
of potential outcomes. This homogeneity hypothesis also aligns
with the decision-making process of risk-sensitive individuals, who, when
faced with two possible interventions that possess identical expected
effects given their baseline characteristics, will sensibly select the one producing
the least variable results. Even when treatment effect modifiers are known to
exist but are unavailable and the second-moment homogeneity hypothesis holds,
this knowledge about unmeasured effect modifiers has no practical
clinical utility.

Importantly, just as with the sharp homogeneous treatment effect hypothesis,
$\Psi_{F}(P_{X,0}) \neq 0$ implies that the second-moment homogeneous
treatment effect hypothesis is violated. And, again,
$\Psi_{F}(P_{X,0})$ can be used to detect departures from this relaxed 
definition of homogeneity even when $V = \emptyset$.

\begin{theorem}\label{thm:constant-ite-violations-abs-diff-var}
  Let $X$, $f^{(a)}(W)$, $\epsilon^{(a)}$, and $\gamma$ be defined as above. If
  $\Psi_F(P_{X,0}) \neq 0$, then $f^{(1)}(W) \neq f^{(0)}(W) + \gamma$ or
  $\mathbb{E}_{P_{X,0}}[\mathbb{V}_{P_{X,0}}[\epsilon^{(1)}|W]] \neq \mathbb{E}_{P_{X,0}}[\mathbb{V}_{P_{X,0}}[\epsilon^{(0)}|W]]$,
  or both. The treatment effect is not second-moment homogeneous.
\end{theorem}

$\Psi_{F}(P_{X,0})$ is therefore a suitable target of causal inference for
guiding, among other things, treatment decisions, health policy recommendations,
and clinical guideline adjustments when the effect modifiers, if any, are
possibly missing from the data considered for decision-making.
Its reliance on the full-data DGP aside, this differential variance estimand can
be used to construct hypothesis tests that detect actionable departures from a practical
homogeneous treatment effect hypothesis. Inference about $\Psi_{F}(P_{X,0})$ is arguably
relevant even when suspected effect modifiers are contained in the data: initial
tests about it can inform whether a more time-consuming, resource-intensive
analysis of the CATE is worth pursuing. What is more, $\Psi_{F}(P_{X,0})$ is
straightforward to interpret, while the CATE can be a possibly opaque functional that
is difficult to communicate to non-statisticians.

Most relevant to our motivating application, however, is that this contrast of the
potential outcome variances provides a natural sensitivity analysis estimand that
complements CATE- and TEM-VIP-based heterogeneous treatment effect analyses.
Inference about $\Psi_{F}(P_{X,0})$ can provide evidence against a second-moment 
homogeneous treatment effect hypothesis following null CATE and TEM-VIP results
--- meaning there is insufficient evidence to reject
$f^{(1)}(W) = f^{(0)}(W) + \gamma$ --- in turn suggesting that the treatment effect
is heterogeneous but the effect modifiers are unmeasured. Indeed, when it can be
assumed that $f^{(1)}(W) = f^{(0)}(W) + \gamma$, testing $\Psi_{F}(P_{X,0})=0$ is
equivalent to testing
$\mathbb{E}_{P_{X,0}}[\mathbb{V}_{P_{X,0}}[\epsilon^{(1)}|W]] = \mathbb{E}_{P_{X,0}}[\mathbb{V}_{P_{X,0}}[\epsilon^{(0)}|W]]$.
A potentially detrimental decision based solely on such a CATE
analysis, like sanctioning the use of a novel, non-inferior intervention that
produces needlessly variable outcomes relative to the prior standard of care, can
thus be avoided by factoring this estimand into decision-making.

We caution that $\Psi_F(P_{X,0}) = 0$ does not generally imply that the
treatment effect is second-moment homogeneous. Noting that
$\mathbb{V}_{P_{X,0}}[Y^{(a)}] = \mathbb{V}_{P_{X,0}}[f^{(a)}(W)] + \mathbb{E}_{P_{X,0}}[\mathbb{V}_{P_{X,0}}[\epsilon^{(a)}|W]]$
by the law of total variance, we may find that
$\mathbb{V}_{P_{X,0}}[f^{(0)}(W)] + \mathbb{E}_{P_{X,0}}[\mathbb{V}_{P_{X,0}}[\epsilon^{(0)}|W]] = \mathbb{V}_{P_{X,0}}[f^{(1)}(W)] + \mathbb{E}_{P_{X,0}}[\mathbb{V}_{P_{X,0}}[\epsilon^{(1)}|W]]$
even when $f^{(1)}(W) \neq f^{(0)}(W) + \gamma$ and
$\mathbb{E}_{P_{X,0}}[\mathbb{V}_{P_{X,0}}[\epsilon^{(1)}|W]] \neq \mathbb{E}_{P_{X,0}}[\mathbb{V}_{P_{X,0}}[\epsilon^{(0)}|W]]$,
though we expect such cancellations to be uncommon. Additionally, relying on this
null hypothesis to detect heterogeneity inherently assumes that distributional
differences between $Y^{(0)}$ and $Y^{(1)}$ are reflected
in their variances; homogeneity hypotheses placing additional restrictions
on the higher moments of the potential outcomes distributions are required
otherwise.

We highlight that contrasts of the potential outcome variances need not be
considered exclusively on the absolute scale. For instance, when it is of interest to
interpret differential variance on the relative scale and it can be assumed that $\sigma^2_F(P_{X,0}; 0)>0$,
we may instead consider the following causal estimand, which we call the
\textit{relative differential variance}:
\begin{equation*}
    \Lambda_F(P_{X,0}) = \frac{\sigma^2_F(P_{X,0}; 1)}{\sigma^2_F(P_{X,0}; 0)} \;.
\end{equation*}
Values larger than one suggest that the potential outcomes associated with the
treatment group have larger variances than those of the control group, whereas
values smaller than one indicate the opposite. Similar to the result about
$\Psi_F(P_{X,0})$ given in
Theorem~\ref{thm:constant-ite-violations-abs-diff-var},
$\Lambda_F(P_{X,0}) \neq 1$ implies that the second-moment homogeneous treatment
effect hypothesis is violated by
Corollary~\ref{cor:constant-ite-violations-rel-diff-var}.

\begin{corollary}\label{cor:constant-ite-violations-rel-diff-var}
  Consider the setting of
  Theorem~\ref{thm:constant-ite-violations-abs-diff-var}. It follows that if
  $\Lambda_F(P_{X,0}) \neq 1$, then $f^{(1)}(W) \neq f^{(0)}(W) + \gamma$ or
  $\mathbb{E}_{P_{X,0}}[\mathbb{V}_{P_{X,0}}[\epsilon^{(1)}|W]] \neq \mathbb{E}_{P_{X,0}}[\mathbb{V}_{P_{X,0}}[\epsilon^{(0)}|W]]$,
  or both. The treatment effect is not second-moment homogeneous.
\end{corollary}

\subsection{Causal Identifiability Conditions}

Now, as previously mentioned, realizations of the full-data DGP are unobservable
owing to the fundamental problem of causal inference
\citep{hollandStatisticsCausalInference1986}. Instead, we have access to realizations
of $n$ i.i.d. random
observations $O = (W, A, Y) \sim P_0 \in \mathcal{M}$. $W$ and $A$ are defined
as in $X$, and $Y$ is a continuous random variable representing the observed
outcome. $P_0$ is the true but unknown observed-data DGP and is fully specified
by $P_{X,0}$ up to the coarsening of the potential outcomes. Analogously,
$\mathcal{M}$ is the nonparametric model of possible observed-data DGPs.

As a first step in performing inference about $\Psi_F(P_{X,0})$ and
$\Lambda_F(P_{X,0})$, we must establish the identifiability of these estimands
in $\mathcal{M}$. We first provide identification conditions for the raw moments
of potential outcomes. Let
$\bar{Q}_{P}^d(W, a) = \mathbb{E}_{P}[Y^d|W,A=a]$ and
$\mu^d(P; a) = \mathbb{E}_{P}[\bar{Q}_P^d(W, a)]$ for $d \in \mathbb{N}$
and $a \in \mathcal{A}$. Again, we omit the superscript when $d=1$ to simplify
notation where possible.

\begin{theorem}\label{thm:moment-identifiability}
  If $A \perp\!\!\!\perp Y^{(a)} | W$ for $a \in\mathcal{A}$
  (conditional exchangeability), $Y = AY^{(1)}+(1-A)Y^{(0)}$ (consistency), and
  $0 < \mathbb{P}_{P_{X,0}}[A = 1 | W] < 1$ a.s. (positivity), then
  $\mu^d_F(P_{X,0}; a) \equiv \mu^d(P_{0}; a)$.
\end{theorem}

Then Corollary~\ref{cor:var-identifiability} follows directly from
Theorem~\ref{thm:moment-identifiability}.

\begin{corollary}\label{cor:var-identifiability}
  Under the conditions of Theorem~\ref{thm:moment-identifiability}, we find
  that, for $a \in \mathcal{A}$,
\begin{equation*}
    \begin{split}
        \sigma^2_F(P_{X,0}; a)
        & = \mu^2_F(P_{X,0}; a) - \mu_F(P_{X,0}; a)^2 \\
        & \equiv \mu^2(P_{0}; a) - \mu(P_{0}; a)^2 \\
        & = \sigma^2(P_{0}; a) \; .
    \end{split}
  \end{equation*}
  Then absolute and relative differential variance estimands are similarly estimable from the observed-data DGPs:
  \begin{equation*}
    \Psi_F(P_{X,0}) \equiv \Psi(P_{0}) = \sigma(P_{0}; 1) - \sigma(P_{0}; 0)
  \end{equation*}
  where $\sigma(P_{0}; a) = \sqrt{\sigma^{2}(P_{0}; a)}$, and
  \begin{equation*}
    \Lambda_F(P_{X,0}) \equiv \Lambda(P_{0}) = \frac{\sigma^2(P_{0}; 1)}{\sigma^2(P_{0}; 0)}\; .
  \end{equation*}
\end{corollary}

The conditional exchangeability, consistency, and positivity assumptions of
Theorem~\ref{thm:moment-identifiability} are standard in the causal inference
literature \citep{hernanCausalInferenceWhat2025}. Conditional exchangeability requires that
distributional parameters of the potential outcomes, like their (centered)
higher moments, are independent of the treatment assignment conditional on
covariates. This implies that there must not be any unmeasured confounding;
$U \subseteq W$ must contain all treatment--outcome confounders. Consistency
requires that the observed outcome corresponds to the potential outcome
associated with the observed treatment condition. Finally, positivity states
that all observations have a non-zero probability of receiving the treatment or
control intervention.

Now, Corollary~\ref{cor:var-identifiability} states that $\Psi(P_{0})$ and
$\Lambda(P_{0})$ have causal interpretations when the conditions of
Theorem~\ref{thm:moment-identifiability} are met. Coupled with
Theorem~\ref{thm:constant-ite-violations-abs-diff-var} and
Corollary~\ref{cor:constant-ite-violations-rel-diff-var}, these estimands can
detect departures from the second-moment homogeneous treatment effect hypothesis
when all effect modifiers are missing from the data so long as the conditional
exchangeability requirement is satisfied. These scenarios arise naturally, for
instance, when the baseline covariate driving the heterogeneous response to
treatment is unknown, like an as-yet unclassified disease subtype attributable
to an undocumented genetic mutation, such that the unmeasured effect modifier is
not expected to confound the treatment--outcome relationship. Further, inference
about $\Psi(P_{0})$ and $\Lambda(P_{0})$ can uncover heterogeneity in RCTs with
complete randomization schemes, like the motivating trials considered here,
without needing to condition on \textit{any} pre-treatment covariates. Recent
work demonstrates that doing so is not only inefficient asymptotically when $W$
contains prognostic variables of the outcome, however, but is also generally
inefficient in finite samples \citep{zaki2026}. This aligns with comparisons
of other marginal estimands' adjusted and unadjusted estimators in RCTs
\citep[see, for example, ][]{benkeserImprovingPrecisionPower2021}.

\section{Inference in a Nonparametric Model}\label{sec:inference}

\subsection{Deriving the EIF of the Potential Outcome Variances}

Having specified conditions under which $\Psi_F(P_{0})$ and
$\Lambda_F(P_{0})$ are estimable in the observed-data DGP, we turn our attention
to performing inference about these estimands. We rely on the general framework
of EIF-based inference in nonparametric models to derive causal machine learning
estimators of $\sigma^2(P_{0}; a)$ and study their asymptotic properties
\citep{vanderlaanTargetedLearningCausal2011, vanderlaanTargetedLearningData2018,
  chernozhukovGenericMachineLearning2018,
  hinesDemystifyingStatisticalLearning2022, kennedyOptimalDoublyRobust2023}. We
then use these results and the functional delta method
\citep{vanderlaanTargetedLearningCausal2011} to develop semiparametric efficient
estimators of $\Psi(P_{0})$ and $\Lambda(P_{0})$.

Denoting the propensity score by
$g_P(W) = \mathbb{P}_P (A=1|W)$, we begin by deriving the EIF of
$\sigma^2(P; a)$, $D^{\sigma^2(a)}(P; O)$, for $P \in \mathcal{M}$.

\begin{proposition}\label{prop:sigma-eif}
  The EIF of $\sigma^2(P; a)$ for some $P \in \mathcal{M}$ is given by
    \begin{equation*}
      \begin{split}
          D^{\sigma^2(a)}(P; O) & =
                                  \frac{I(A=a)\left(Y^2 - \bar{Q}_P^2(W,a) + 2\mu(P; a) \left(\bar{Q}_P(W,a)- Y\right)\right)}{I(A=1)g_P(W) + I(A=0)(1-g_P(W))} + \\
                                & \qquad \qquad\bar{Q}_P^2(W,a)  - 2\bar{Q}_P(W,a)\mu(P; a) + \mu(P; a)^2 - \sigma^2(P; a)  \;.
      \end{split}
    \end{equation*}
\end{proposition}

It is straightforward to show that the variance of $D^{\sigma^2(a)}(P;O)$ is
bounded under the assumption of positivity and that
$\mathbb{E}_{P_0}[Y^4] < \infty$. Then $\sigma^2(P;a)$ is a pathwise differentiable
parameter in $\mathcal{M}$ for which regular and asymptotically linear estimators
can be constructed \citep{vanderlaanTargetedLearningCausal2011}.

\subsection{Estimators of the Potential Outcome Variances}\label{sec:estimators}

We now present several nonparametric estimators of $\sigma^2(P_0; a)$. Let $g_n$,
$\bar{Q}_n^d$, and $\mu_n(a)$ be estimators of $g_{P_0}$, $\bar{Q}_{P_0}^d$, and
$\mu(P_0; a)$, respectively, and included in $\hat{P}_n$, the empirical
distribution $P_{n}$ augmented with nuisance estimators. Let us further assume that the propensity score estimator is selected such that $0<g_n(W)<1$ a.s. Then the one-step estimator \citep{pfanzaglCONTRIBUTIONSGENERALASYMPTOTIC1985,
  bickelEfficientAdaptiveEstimation1998} is defined as:
\begin{equation*}
    \begin{split}
        \sigma^2_{\text{OS}}(\hat{P}_n;a)
        & = \sigma^2(\hat{P}_n ;a) + \mathbb{E}_{P_n}\left[D^{\sigma^2(a)}(\hat{P}_n; O)\right]\\
        & = \mathbb{E}_{P_n}\bigg[\frac{I(A=a)\left(Y^2 - \bar{Q}_n^2(W,a) + 2\mu_n(a) \left(\bar{Q}_n(W,a)- Y\right)\right)}{I(A=1)g_n(W) + I(A=0)(1-g_n(W))} + \\
        & \qquad \qquad \qquad \bar{Q}_n^2(W, a) - 2\bar{Q}_n(W, a)\mu_n(a)\bigg] + \mu_n(a)^2 \;.
    \end{split}
\end{equation*}

We note that $\sigma^2_{\text{OS}}(\hat{P}_n;a)$ is equivalent to
the estimating equation estimator \citep{vanderlaanUnifiedMethodsCensored2003,
  chernozhukovDoubleDebiasedNeyman2017} of $\sigma^2(P_0;a)$
because $D^{\sigma^2}(P;O)$ is linear in $\sigma^2(P;a)$.

The derivation of a TMLE for $\sigma^2(P_0; a)$ is more involved. First, the
outcome variable is rescaled to the unit interval prior to
estimating the nuisance parameters. Given outcomes $\{Y\}_{i=1}^n$, the scaled
outcome for observation $i$ is computed as
$(Y_i - \min_i\{Y_i\})/(\max_i\{Y_i\}- \min_i\{Y_i\})$. In an abuse of notation,
we avoid re-defining nuisance parameters and estimators to reflect the use of this
transformed outcome to simplify exposition. Then $\hat{P}_n$ is tilted to obtain
$\hat{P}_n^\star$ such that
$\mathbb{E}_{P_n}[D^{\sigma^2(a)}(\hat{P}_n^\star; O)] = 0$. For this
estimand, the asymptotic bias term can be negated by tilting $\bar{Q}_n$ and
$\bar{Q}_n^2$. Define the negative log-likelihood loss function for
$\bar{Q}^d$ as
\begin{equation*}
    L(O, a; \bar{Q}^d) = -\log\left\{\bar{Q}^d(W, a)^{Y^d}\left(1 - \bar{Q}^d(W, a)\right)^{(1-Y^d)}\right\}
\end{equation*}
and a parametric working submodel for $\bar{Q}^d$ as
\begin{equation*}
    \bar{Q}^d(\eta)(O,a) = \text{logit}^{-1}\left(\text{logit}\left(\bar{Q}^d(W,a)\right) + \eta H(O,a)\right) \;,
\end{equation*}
where
\begin{equation*}
    H(O, a) = \frac{I(A=a)}{I(A=1)g(W) + I(A=0)(1-g(W))} \;.
\end{equation*}
The tilted nuisance estimators $\bar{Q}_n^\star$ and $\bar{Q}_n^{2\star}$ of
$\bar{Q}_{P_0}$ and $\bar{Q}_{P_0}^2$, respectively, are obtained by computing
$\eta_n^{d\star}$ such that
\begin{equation*}
    \eta^{d\star}_n = \text{arg min}_\eta \mathbb{E}_{P_n}\left[L\left(O, a; \bar{Q}^d_n(\eta)(O, a)\right)\right]
\end{equation*}
for $d \in \{1, 2\}$. Then
$\bar{Q}_n^{d\star} = \bar{Q}^d_n(\eta^{d\star}_n)$, where, again, we let
$\bar{Q}_n^{\star}=\bar{Q}_n^{1\star}$ to lighten notation. We
highlight that, in addition to $\bar{Q}_n^d$, $\bar{Q}_n^{d\star}$ is a function
of $A$, $Y$, and $g_n$ through $H_n(O, a)$ and $\eta_n^{d\star}$. $H_n$ is the
empirical equivalent of $H$ that relies on $g_n$ in place of $g$. Importantly,
computing $\eta^{d\star}_n$ is straightforward: it is the maximum likelihood
estimator of the univariate logistic regression slope coefficient obtained by
regressing $Y^d$ on $H_n$ using $\text{logit}(\bar{Q}_n^d)$ as an offset
\citep{laanTargetedMaximumLikelihood2006}. The tilted plug-in estimator
$\hat{P}^\star_n$ is therefore obtained by replacing $\bar{Q}_n$ and
$\bar{Q}^2_n$ by $\bar{Q}_n^\star$ and $\bar{Q}^{2\star}_n$ in $\hat{P}_n$,
respectively. Owing to the choice of loss function and parametric working
submodels, we find that
$\mathbb{E}_{P_n}[D^{\sigma^2(a)}(\hat{P}_n^\star; O)] = 0$. Finally, the TMLE
of $\sigma^2(P_0; a)$ is given by
\begin{equation*}
    \begin{split}
      \sigma^2_\text{TMLE}(\hat{P}_n; a)
      & = \left(\max_i\{Y_i\}- \min_i\{Y_i\}\right)^2 \; \sigma^2(\hat{P}_n^\star; a) \\
      & = \left(\max_i\{Y_i\}- \min_i\{Y_i\}\right)^2 \; \left(\mathbb{E}_{P_n}\left[\bar{Q}^{2\star}_n(O, a)\right] - \mathbb{E}_{P_n}\left[\bar{Q}^{\star}_n(O, a)\right]^2\right) \;.
    \end{split}
\end{equation*}

We additionally derived cross-fitted \citep{zhengCrossValidatedTargetedMinimumLossBased2011,
  chernozhukovDoubleDebiasedNeyman2017} versions of the one-step estimator and TMLE,
denoted $\sigma^{2}_{\text{CFOS}}(\hat{P}_{n}; a)$ and
$\sigma^{2}_{\text{CFTMLE}}(\hat{P}_{n}; a)$, respectively.

\subsection{Asymptotic Properties of the Potential Outcome Variance Estimators}

We now detail the asymptotic properties of these nonparametric estimators of
$\sigma^2(P_0; a)$, noting that
$\lVert f_n(O) - f(O) \rVert_2^2 = \int (f_n(o) - f(o))^2 dP(o)$ for some
estimator $f_n$ of $f: \mathcal{O} \mapsto \mathbb{R}$ under
$P \in \mathcal{M}$ where $\mathcal{O}$ is the support of $O$.

\begin{theorem}\label{thm:double-robustness}
  \textbf{Double Robustness:} If for $a \in \mathcal{A}$
    \begin{enumerate}
      \item $\lVert g_n(W) - g_{P_0}(W) \rVert_2 = o_P(1)$ or
            $\lVert \bar{Q}_n^d(W, a) - \bar{Q}_{P_0}^d(W, a) \rVert_2 = o_P(1)$
            for $d = 1$ and $d = 2$, and
      \item $\lvert \mu_n(a) - \mu(P_0;a) \rvert = o_P(1)$,
    \end{enumerate}
    then
    $\sigma^2_\text{OS}(\hat{P}_n; a) \overset{P}{\rightarrow} \sigma^2(P_0; a)$.
    The same result holds for $\sigma^2_\text{CFOS}$, $\sigma^2_\text{TMLE}$,
    and $\sigma^2_\text{CFTMLE}$.
\end{theorem}

While the conditions for consistency in Theorem~\ref{thm:double-robustness} may
appear more stringent than traditional double robustness results at first
glance, we highlight that the first condition ensures that the second is
satisfied when $\mu_n(a)$ is a double robust estimator of $\mu(P_0; a)$. That
is, for a double robust estimator $\mu_n(a)$,
$\lvert \mu_n(a) - \mu(P_0;a)\rvert = o_P(1)$ when
$\lVert g_n(W) - g_{P_0}(W) \rVert_2 = o_P(1)$ or
$\lVert \bar{Q}_n(W, a) - \bar{Q}_{P_0}(W, a) \rVert_2 = o_P(1)$. This implies
that the previously introduced estimators of $\sigma^2(P_0; a)$ are consistent
in RCTs like the TTM and TTM2 trials where the propensity score is known, regardless
of whether the estimators of $\bar{Q}_{P_0}^d(W, a)$ are consistent. In observational studies,
these consistency conditions may be satisfied by estimating $\mu(P_0;a)$ with a
double robust estimator and $g_{P_0}$, $\bar{Q}_{P_0}$, and $\bar{Q}_{P_0}^2$
with flexible machine learning techniques like the Super Learner ensemble method
\citep{laanSuperLearner2007}.

We next consider the conditions required of these estimators to be
asymptotically linear and efficient. We present these results with respect to the
non-cross-fitted estimators, though they are straightforwardly extended.

\begin{theorem}\label{thm:asymptotic-linearity}
  \textbf{Asymptotic Linearity}: If for $a \in \mathcal{A}$
    \begin{enumerate}
      \item There exists a $P_0$-Donsker class $\mathbb{G}_0$ such that
            $\mathbb{P}_{P_0}[D^{\sigma^2(a)}(\hat{P}_n; O) \in \mathbb{G}_0] \rightarrow 1$,
      \item $\lVert D^{\sigma^2(a)}(\hat{P}_n; O)~-~D^{\sigma^2(a)}(P_0; O)\rVert_2~=~o_P(1)$,
      \item
            $\lVert g_{n}(W) - g_{P_0}(W) \rVert_2 \lVert \bar{Q}_{n}(W, a) - \bar{Q}_{P_0}(W, a) \rVert_2 = o_P(n^{-1/2})$,
      \item
            $\lVert g_{n}(W) - g_{P_0}(W) \rVert_2 \lVert \bar{Q}^2_{n}(W, a) - \bar{Q}^2_{P_0}(W, a) \rVert_2 = o_P(n^{-1/2})$,
            and
      \item $\lvert \mu_{n}(a) - \mu(P_0; a) \rvert = o_P(n^{-1/4})$,
    \end{enumerate}
    then
    $\sigma^2_\text{OS}(\hat{P}_n; a) - \sigma^2(P_0; a) = \mathbb{E}_{P_n}[D^{\sigma^2(a)}(P_0; O)] + o_P(n^{-1/2})$.
    The same result holds for $\sigma^2_\text{TMLE}$. $\sigma^2_\text{CFOS}$ and
    $\sigma^2_\text{CFTMLE}$ are asymptotically linear under conditions 2, 3, 4, and 5.
\end{theorem}

The first condition of Theorem~\ref{thm:asymptotic-linearity} is an unverifiable
assumption that is satisfied when the EIF is Donsker. This is equivalent to
placing weak regularity conditions on the nuisance parameters: they must be
c\`{a}dl\`{a}g (\textit{continue \`{a} droite, limite \`{a} gauche}: right
continuous with left limits) and have finite supremum and sectional variation
norms \citep{Gill1995}. Note that a class $\mathcal{F}$ with bounded supremum
norm is $P$-Donsker if $\mathcal{F}$ is pre-Gaussian and the empirical process
$\mathbb{G}(\mathcal{F}) = \{ \mathbb{G}(f), f \in \mathcal{F} \}$ converges
weakly under $L_{\infty}(P)$ to the Gaussian process $\mathbb{G}_P(\mathcal{F})$
\citep{bickelEfficientAdaptiveEstimation1998,
  vanderlaanTargetedLearningCausal2011}. Of note, this assumption can be avoided
by using cross-fitted estimators
\citep{zhengCrossValidatedTargetedMinimumLossBased2011}. Their reliance on
sample-splitting may reduce their finite-sample efficiency relative to their
non-cross-fitted counterparts when the Donsker condition is in fact satisfied,
however.

The second condition, along with the third and fourth double rate robustness
conditions, is satisfied when, for example,
$\lVert g_{n}(W) - g_{P_0}(W) \rVert_2 = o_P(n^{-1/4})$,
$\lVert \bar{Q}_{n}(W, a) - \bar{Q}_{P_0}(W, a) \rVert_2 = o_P(n^{-1/4})$, and
$\lVert \bar{Q}^2_{n}(W, a) - \bar{Q}^2_{P_0}(W, a) \rVert_2 = o_P(n^{-1/4})$.
Of course, other nuisance estimator convergence rates satisfy these conditions as well.

Finally, the last condition of Theorem~\ref{thm:asymptotic-linearity} is
immediately satisfied when the third condition of the theorem is fulfilled and
$\mu_{n}$ is a double robust estimator. This is because double robust estimators
of $\mu(P_0; a)$ are efficient and asymptotically linear under the
aforementioned double rate robustness requirement, converging at rates of
$O_P(n^{-1/2})$. In fact, the contribution of the double robust estimation of
$\mu(P_0; a)$ to the second order remainder terms of considered estimators is
$O_P(n^{-1})$. As such, in large enough samples, finite-sample error incurred
when estimating $\mu(P_0; a)$ has negligible impact on the finite-sample errors
of these estimators when using double robust estimators of this nuisance
parameter.

Theorem~\ref{thm:asymptotic-linearity} implies that
$\sqrt{n}(\sigma^2_\text{OS}(a)-\sigma^2(P_0;a))~\overset{D}{\rightarrow}~N(0, \mathbb{E}_{P_0}[D^{\sigma^2(a)}(P_0; O)^2])$
when the previously described conditions are satisfied. The same is true for
$\sigma^2_\text{TMLE}$, $\sigma^2_\text{CFOS}$, and $\sigma^2_\text{CFTMLE}$.
This result can be used to straightforwardly construct $\alpha$-level Wald-type
confidence intervals \citep{vanderlaanTargetedLearningCausal2011}, permitting hypothesis
testing about $\sigma^2(P_0; a)$. This result also has important implications for
potential outcome variance inference in RCTs: since $g_{P_0}$ is known in randomized DGPs,
these causal machine learning estimators are efficient and asymptotically normal with a
variance that can be consistently estimated when $\mu_{n}$ is a double robust estimator,
$\lVert \bar{Q}_{n}(W, a) - \bar{Q}_{P_0}(W, a) \rVert_2 = o_P(1)$, and
$\lVert \bar{Q}_{n}^2(W, a) - \bar{Q}_{P_0}^2(W, a) \rVert_2 = o_P(1)$. 
These estimators remain asymptotically linear even when the nuisance estimators
of $\bar{Q}_{P_0}$ and $\bar{Q}^2_{P_0}$ are misspecified, although they are no
longer efficient. Their asymptotic variance is instead equal to
$\mathbb{E}_{P_0}[\tilde{D}^{\sigma^2(a)}(P_0; O)^2]/n$, where
\begin{equation*}
  \begin{split}
      \tilde{D}^{\sigma^2(a)}(P_0; O) & =
                              \frac{I(A=a)\left(Y^2 - \tilde{\bar{Q}}_{P_0}^2(W,a) + 2\mu(P_0; a) \left(\tilde{\bar{Q}}_{P_0}(W,a)- Y\right)\right)}{I(A=1)g_{P_0}(W) + I(A=0)(1-g_{P_0}(W))} + \\
                            & \qquad \qquad\tilde{\bar{Q}}_{P_0}^2(W,a)  - 2\tilde{\bar{Q}}_{P_0}(W,a)\mu(P_0; a) + \mu(P_0; a)^2 - \sigma^2(P_0; a)  \;.
  \end{split}
\end{equation*}
Here, $\tilde{\bar{Q}}_{P_0}$ and $\tilde{\bar{Q}}_{P_0}^2$ are the limiting
quantities of misspecified $\bar{Q}_n$ and $\bar{Q}_n^2$, respectively.

\subsection{Differential Variance Inference}\label{subsec:diff-var}

We now propose causal machine learning estimators of $\Psi(P_0)$ and
$\Lambda(P_0)$ in nonparametric models based on the causal machine learning
estimators of $\sigma^{2}(P_{0}; a)$. It is straightforward to define the
one-step estimator of the absolute differential variance estimand:
\begin{equation*}
  \Psi_\text{OS}(\hat{P}_n) = \sqrt{\sigma^2_\text{OS}(\hat{P}_n; 1)} - \sqrt{\sigma^2_\text{OS}(\hat{P}_n; 0)} \;.
\end{equation*}
The variance of this estimator may be obtained through the EIF of $\Psi(P_0)$,
which itself is based on the EIF of $\sigma^{2}(P_{0};a)$ derived in
Proposition~\ref{prop:sigma-eif} and the functional delta method. The EIF of
$\Psi(P_0)$ is given by:
\begin{equation*}
  D^{\Psi}(P; O) = \frac{D^{\sigma^2(1)}(P; O)}{2\sqrt{\sigma^2(P;1)}} - \frac{D^{\sigma^2(0)}(P; O)}{2\sqrt{\sigma^2(P;0)}} \;.
\end{equation*}

The cross-fitted one-step estimator, $\Psi_\text{CFOS}$, is similarly defined,
as are the TMLE and the cross-fitted TMLE estimators, $\Psi_\text{TMLE}$ and
$\Psi_\text{CFTMLE}$. The latter estimators target $\sigma^{2}(P_{0}; 0)$ and
$\sigma^{2}(P_{0}; 1)$ separately. All estimators are double robust and
asymptotically linear under the conditions provided in
Theorem~\ref{thm:double-robustness} and Theorem~\ref{thm:asymptotic-linearity},
respectively. As such,
$\sqrt{n}(\Psi_\text{OS}-\Psi(P_{0})) \overset{D}{\rightarrow} N(0, \mathbb{E}_{P_0}[D^\Psi(P_0; O)^2])$.
The same is true for $\Psi_\text{CFOS}$, $\Psi_\text{TMLE}$, and
$\Psi_\text{CFTMLE}$. These efficient estimators can therefore be used to perform tests of
the second-moment homogeneous treatment effect hypothesis all while providing
asymptotic Type-I error rate guarantees in a nonparametric model where effect
modifiers are missing. Additionally, in RCTs, these estimators can be used to
construct valid hypothesis tests under misspecified $\bar{Q}_n$ and $\bar{Q}^2_n$.

As for the relative differential variance estimand $\Lambda(P_0)$, its one-step
estimator is given by:
\begin{equation*}
  \Lambda_\text{OS}(\hat{P}_n) = \frac{\sigma^2_\text{OS}(\hat{P}_n; 1)}{\sigma^2_\text{OS}(\hat{P}_n; 0)} \;.
\end{equation*}
The TMLE, cross-fitted one-step estimator, and cross-fitted TMLE are similarly
derived, and denoted by $\Lambda_\text{CFOS}$, $\Lambda_\text{TMLE}$, and
$\Lambda_\text{CFTMLE}$, respectively.

Again relying on Proposition~\ref{prop:sigma-eif} and the functional delta
method, we may derive the variance of this estimator based on the EIF of
$\Lambda(P_0)$:
\begin{equation*}
  D^{\Lambda}(P; O) = \frac{D^{\sigma^2(1)}(P; O)}{\sigma^2(P; 0)} - \frac{\sigma^2(P; 1)D^{\sigma^2(0)}(P; O)}{\sigma^2(P; 0)^2} \;.
\end{equation*}
As with the estimators of $\Psi(P_0)$, the proposed nonparametric estimators of
$\Lambda(P_0)$ are consistent under the conditions provided in
Theorem~\ref{thm:double-robustness} and are asymptotically linear and efficient
under the conditions of Theorem~\ref{thm:asymptotic-linearity}.

\section{Simulations}\label{sec:simulations}

We assess the proposed estimators' performance through three simulation studies.
Simulation Study 1 investigates the double robustness of the relative
differential variance estimators established in
Theorem~\ref{thm:double-robustness}. Simulation Study 2 evaluates the asymptotic
linearity of the absolute differential variance estimators established in
Theorem~\ref{thm:asymptotic-linearity}. Simulation Study 3 gauges the absolute
differential variance estimators' ability to uncover violations of the
second-moment homogeneous treatment effect hypothesis in RCTs. In all numerical
experiments, $500$ datasets are generated at each of the following sample sizes:
$n \in \{125, 250, 500, 1000, 2000\}$.

\subsection{Simulation Study 1: Double Robustness}

We consider a DGP representative of an observational study:
\begin{equation*}
  \begin{split}
    W_1 &\sim \text{Bern}(0.3), \\
    W_2 &\sim N(0,1), \\
    A \mid W_1, W_2 &\sim \text{Bern}(\text{logit}^{-1}((1 + W_1 + W_2)/4)), \\
    Y \mid A, W_1, W_2 &\sim N(1 + A + W_1 + W_2 + AW_2 + W_1 W_2, 1 + A)
  \end{split}
\end{equation*}
Here, $W_1$ and $W_2$ are the independent baseline covariates, $A$ is the binary
treatment variable, and $Y$ is the outcome variable. To introduce treatment
effect heterogeneity, we set the variance of $Y|W_1, W_2, A$ to depend on $A$
and specify that $\mathbb{E}_{P_0}[Y|W_1, W_2, A]$ includes an interaction
between $W_2$ and $A$. That is, $W_2 \in V$ is a treatment effect modifier.

We take as target of inference the relative differential variance,
$\Lambda(P_0) = 2.47$, and evaluate the absolute empirical bias of
$\Lambda_\text{OS}$, $\Lambda_\text{TMLE}$, $\Lambda_\text{CFOS}$, and
$\Lambda_\text{CFTMLE}$ under four nuisance parameter estimation scenarios. In
the first scenario, all nuisance parameter estimators are well-specified. In the
second scenario, the propensity score estimator is well-specified and the
outcome regression estimators are misspecified. In the third scenario, the
propensity score estimators are misspecified and the outcome regression
estimators are well-specified. In the final scenario, all nuisance parameters
are misspecified.

A main-terms logistic regression conditioning on all confounders is used to
estimate the $g_{P_0}$ in scenarios where the propensity score estimator is
well-specified. When the propensity score estimator is misspecified, a logistic
regression conditioning on $W_2$ is used instead. Random Forest estimators are
used to estimate $\bar{Q}_{P_0}$ and $\bar{Q}_{P_0}^2$ in scenarios where the
outcome regressions are well-specified \citep{breimanRandomForests2001}. When
the outcome regressions are misspecified, these nuisance parameters are
estimated by the treatment groups' mean outcomes and mean squared outcomes,
respectively.

\begin{figure}
    \centering
    \includegraphics[width=1.0\textwidth]{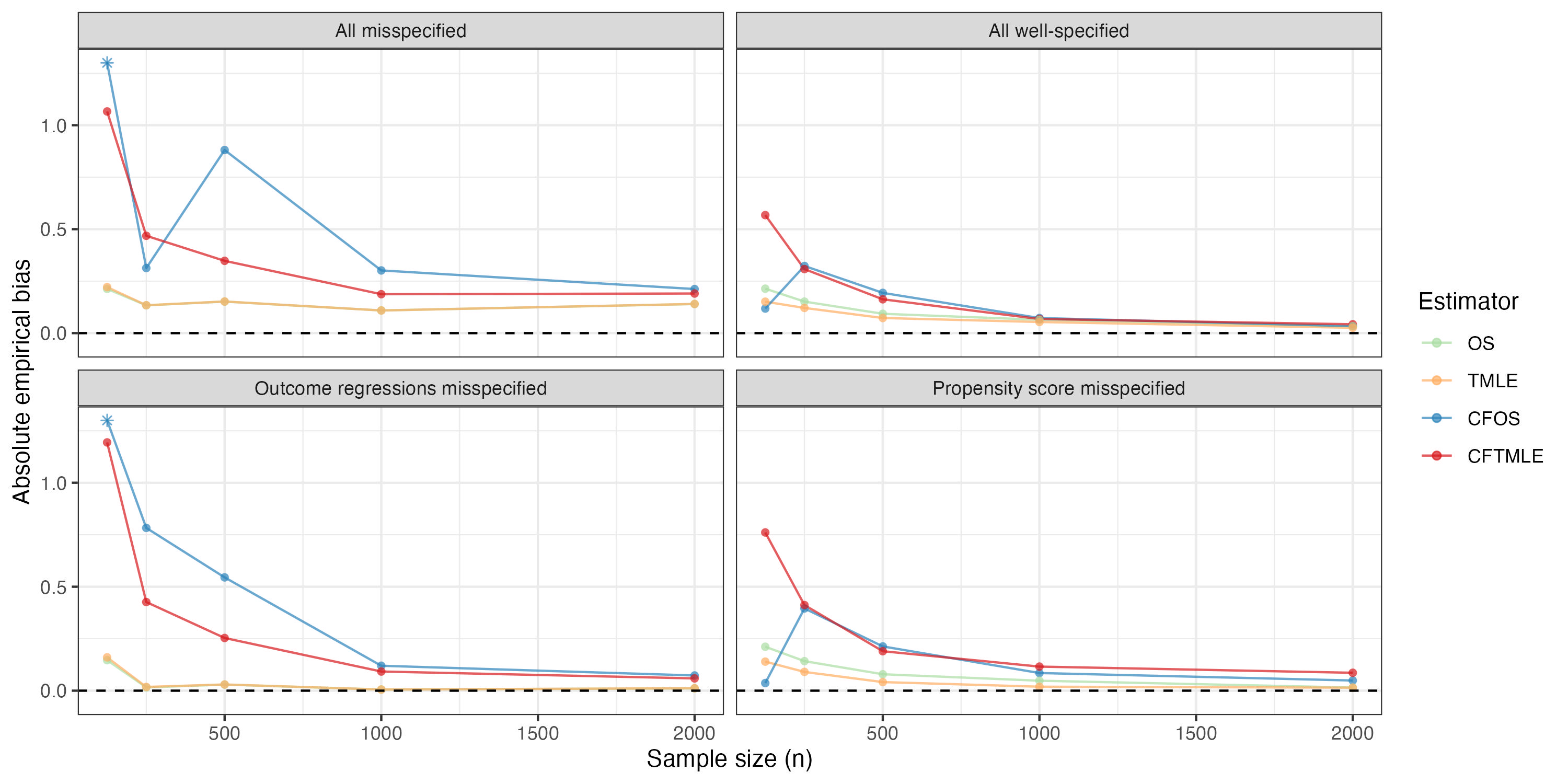}
    \caption{\textbf{Double robustness:} The absolute empirical bias of the
      relative differential variance one-step estimator and TMLE, as well as
      their cross-fitted counterparts, across various sample sizes and nuisance
      parameter estimator scenarios. Outlying values are truncated and
      represented by stars to facilitate the interpretation of results. The
      dashed horizontal lines correspond to the idealized absolute empirical
      bias.}
    \label{fig:double_robustness}
\end{figure}

The results of this simulation study are reported in
Figure~\ref{fig:double_robustness}. As expected, the estimators' absolute
empirical biases are negligible as sample size increases in all but the scenario
where every nuisance parameter estimator is misspecified.

These results suggest that $\Lambda_\text{OS}$, $\Lambda_\text{TMLE}$,
$\Lambda_\text{CFOS}$, and $\Lambda_\text{CFTMLE}$ are double robust under the
considered DGP. Of note, $\Lambda_\text{OS}$ and $\Lambda_\text{TMLE}$ have
lower empirical bias than $\Lambda_\text{CFOS}$ and $\Lambda_\text{CFTMLE}$
across all scenarios. This difference, starkest in the smaller sample sizes, is
likely induced by the cross-fitted estimators' sample splitting.

\subsection{Simulation Study 2: Asymptotic Linearity}

We consider another DGP resembling that of an observational study:
\begin{equation*}
  \begin{split}
    W_1 &\sim N(0,1), \\
    W_2 &\sim N(0,1), \\
    A \mid W_1, W_2 &\sim \text{Bern}(\text{logit}^{-1}((1 + W_1 + W_2)/4)), \\
    Y \mid A, W_1, W_2 &\sim N(1 - 2A + W_1^2 + 2I(W_2 < 0), 1)
  \end{split}
\end{equation*}
Variables $W_1$, $W_2$, $A$, and $Y$ are interpreted as in the previous
simulation study. Unlike Simulation Study 1, however, this DGP does not exhibit
any heterogeneous treatment effects.

In this simulation study, we aim to perform inference about the absolute
differential variance, $\Psi(P_0) = 0$, using $\Psi_\text{OS}$,
$\Psi_\text{TMLE}$, $\Psi_\text{CFOS}$, and $\Psi_\text{CFTMLE}$. In particular,
we investigate whether these estimators' asymptotic linearity properties are
approximately achieved in finite samples, as evidenced by their empirical bias,
variance, and coverage.

The absolute differential variances' nuisance parameters are estimated using
Super Learner ensembles. The Super Learner library for $g_{P_0}$ is made up of a
main-terms logistic regression, a multivariate quadratic logistic regression, a
multivariate adaptive regression splines estimator
\citep{friedmanMultivariateAdaptiveRegression1991}, and Random Forests. The
Super Learner library for $\bar{Q}_{P_0}$ is comprised of a main-terms linear
model, a multivariate quadratic linear model, a multivariate adaptive regression
splines estimator, and Random Forests. Finally, the Super Learner library for
$\bar{Q}_{P_0}^2$ consists of a multivariate quadratic linear model and Random
Forests.

\begin{figure}
    \centering
    \includegraphics[width=1.0\textwidth]{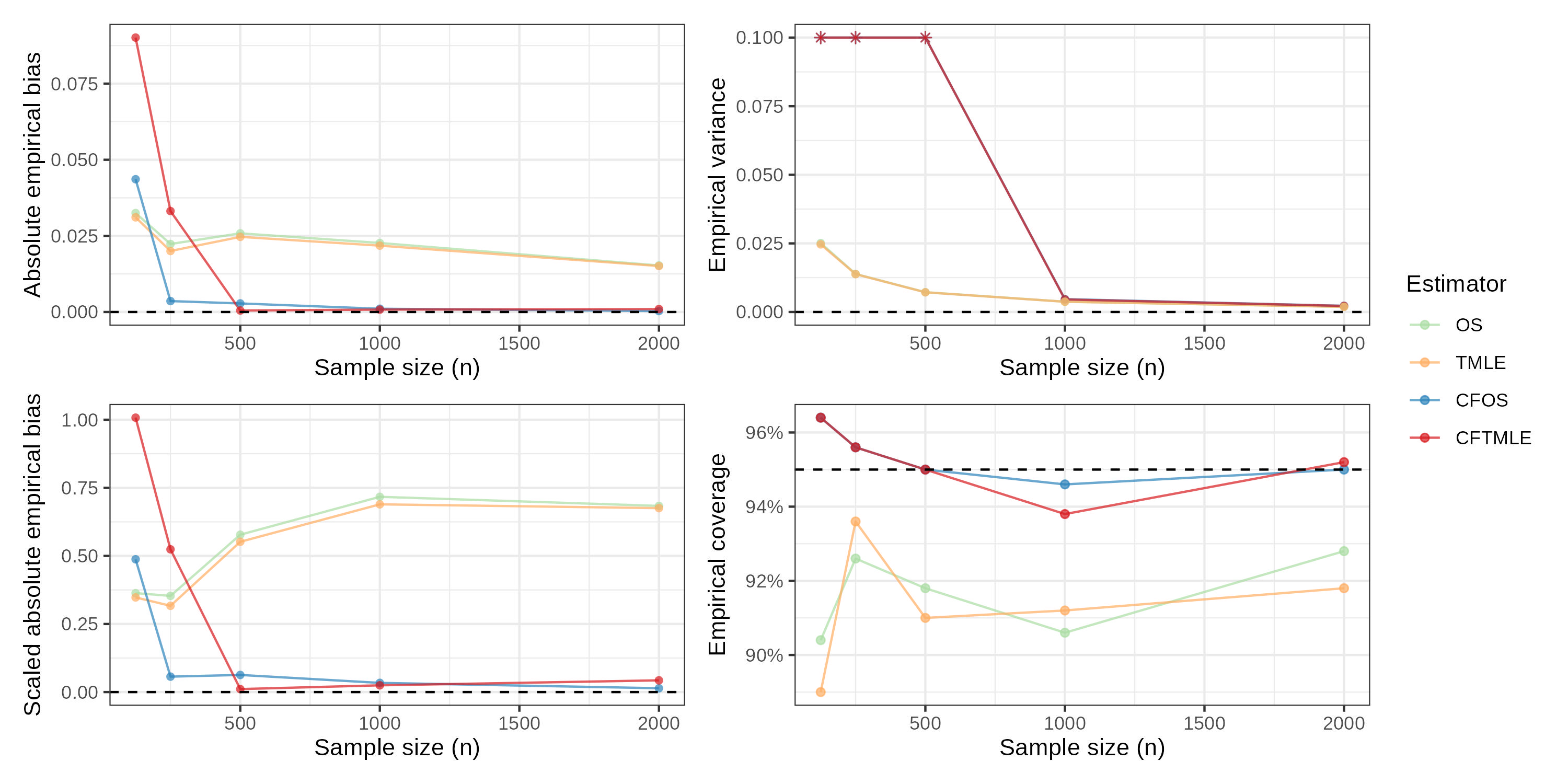}
    \caption{\textbf{Asymptotic linearity:} The absolute empirical bias,
      empirical variance, scaled absolute empirical bias, and empirical coverage
      of the one-step estimator, TMLE, cross-fitted one-step estimator, and
      cross-fitted TMLE for the absolute differential variance. Outlying values
      are truncated and represented by stars to facilitate the interpretation of
      results. The dashed horizontal lines correspond to the idealized values
      for the considered empirical metrics.}
    \label{fig:asymptotic_linearity}
\end{figure}

The results of this experiment are presented in Figure
\ref{fig:asymptotic_linearity}. As expected, absolute empirical bias and
empirical variance become negligible as sample size increases. At smaller sample
sizes, the empirical bias and variance of the cross-fitted estimators are
substantially larger than those of the non-cross-fitted estimators. This behavior is due to the cross-fitted estimators sometimes
producing near-zero potential outcome variance estimates. However, the
cross-fitted estimators outperform the others in terms of bias starting from
$n=500$. The scaled absolute empirical bias, obtained by multiplying the
absolute empirical bias by $\sqrt{n}$, is decreasing for all estimators as of
$n=1000$, suggesting that these estimators are asymptotically root-$n$
consistent. Finally, cross-fitted estimators' coverage achieves the nominal
proportion in samples of size $1000$ and $2000$, while their non-cross-fitted
counterparts produce empirical coverages slightly below --- but converging
towards --- the desired proportion.

The finite-sample performance of the estimators is consistent with asymptotically linearity, though the cross-fitted estimators' performance is markedly better than that of the one-step estimator and TMLE in large samples. Their differential finite-sample behavior might be explained by the use of flexible machine learning algorithms to estimate nonparametric nuisance parameters. Flexible nuisance parameter estimators may be prone to suboptimal bias-variance trade-offs in finite samples \citep{zhengCrossValidatedTargetedMinimumLossBased2011} --- unless sample-splitting strategies are incorporated --- resulting in a slight bias that decreases with growing sample sizes and mildly anti-conservative coverage.

\subsection{Simulation Study 3: Detecting Heterogeneous Treatment Effects}

We consider the following RCT DGP with an effect modifier that is (possibly)
subject to measurement error:

\begin{equation*}
  \begin{split}
    W_1 & \sim N(0, 1) \\
    W_2 & \sim N(0, 1) \\
    W_{3} & \sim \text{Bern}(0.5) \\
    M & \sim \text{Bern}(m) \\
    R & \sim \text{Bern}(0.2)\\
    W_{3}^{\text{obs}} &=
                         \begin{cases}
                           R & \text{if } M = 1 \\
                           W_{3} & \text{if } M = 0
                         \end{cases} \\
    A | W_1, W_2, W_{3} = A & \sim \text{Bern}(0.5) \\
    Y | W_1, W_2, W_{3}, A & \sim N(1 - 2A + W_1^2 + 2I(W_2 < 0) + 4AW_{3}, 1)
  \end{split}
\end{equation*}
Here, $W_1$ and $W_2$ are prognostic variables and $W_{3} \in V$ is a true but
possibly mismeasured effect modifier. The measurement error of $W_{3}$ is
introduced through $M$, a Bernoulli random variable with probability $m$
representing the likelihood of $W_{3}$ being replaced by a Bernoulli random
variable $R$. Thus, $W_{3}^\text{obs}$ is the observed effect modifier subject
to measurement error; $W_{3}$, $R$, and $M$ are unobserved. Binary variable $A$
represents the arm assignment, and $Y$ is the continuous outcome variable.

In this numerical experiment, we aim to detect treatment effect heterogeneity in
four different effect modifier measurement error scenarios:
$m \in \{0, 1/3, 2/3, 1\}$. We accomplish this by performing hypothesis tests
about the second-moment homogeneous treatment effect hypothesis using the
absolute differential variance estimators $\Psi_\text{OS}$, $\Psi_\text{TMLE}$,
$\Psi_\text{CFOS}$, and $\Psi_\text{CFTMLE}$. These estimators' nuisance
parameters are estimated with (misspecified) main-terms linear models,
reflecting the standard practice for covariate adjustment in RCTs
\citep{fdaAdjustingCovariatesRandomized2024}, with the exception of the
propensity scores. These differential variance estimators use the known
propensity scores instead. We then benchmark these tests against the treatment
effect heterogeneity omnibus test of
\citet{chernozhukovGenericMachineLearning2018} implemented in the \textit{grf} R
package for the causal forest CATE estimator \citep{grf}, comparing their
empirical power at a significance cutoff of $0.05$.

\begin{figure}
    \centering
    \includegraphics[width=1.0\textwidth]{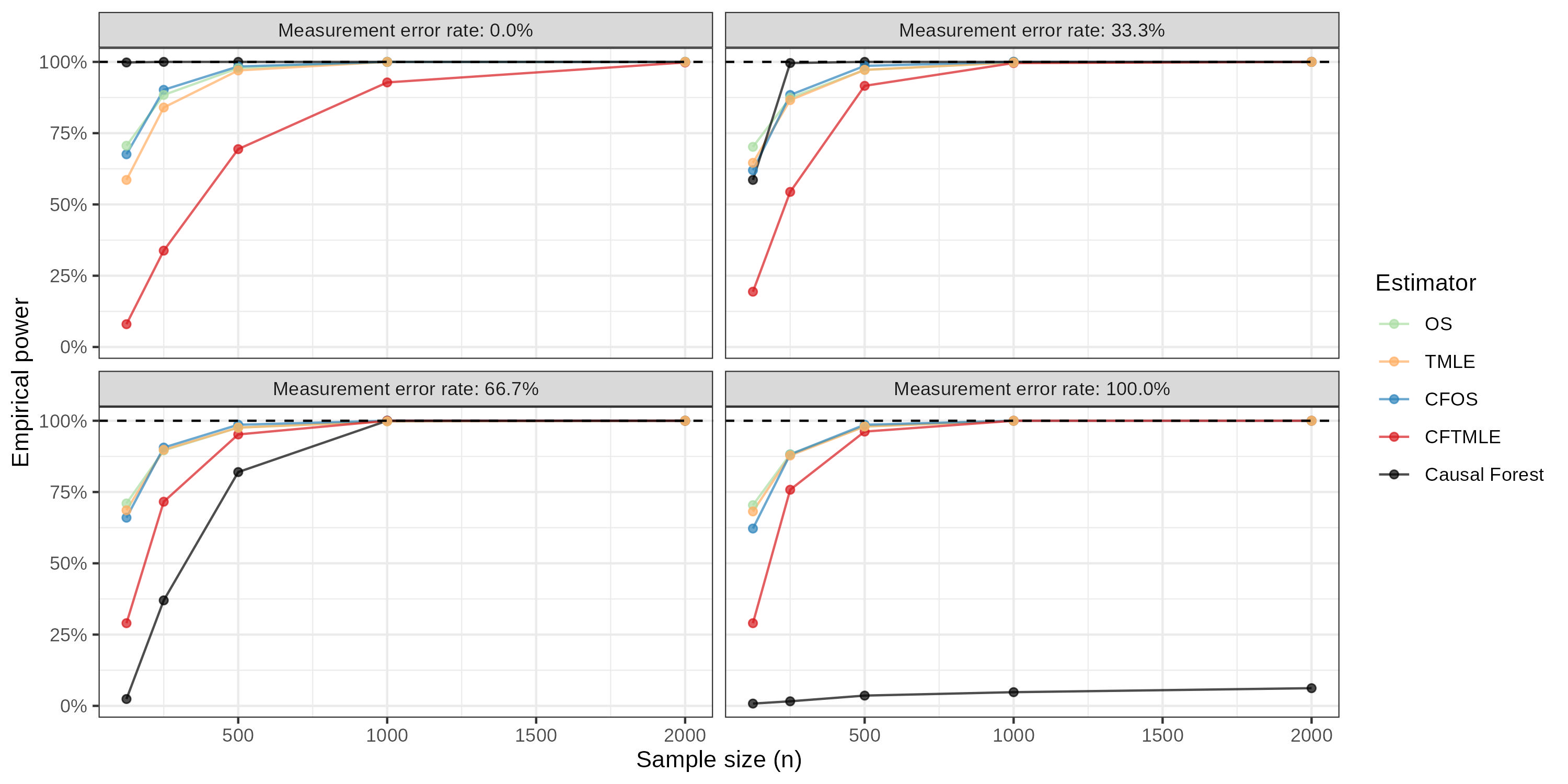}
    \caption{\textbf{Homogeneous Treatment Effect Hypothesis Testing:} The
      empirical power of homogeneous treatment effect hypothesis tests based on
      the absolute differential variance estimators and causal forests in
      several treatment effect modifier mismeasurement scenarios. The dashed
      horizontal lines correspond to the idealized empirical power.}
    \label{fig:constant_ITE}
\end{figure}

The empirical power of the hypothesis tests based on the absolute differential
variance estimators and the causal forest are presented in
Figure~\ref{fig:constant_ITE}. In the absence of measurement error, the causal
forest approach uniformly dominates the differential variance estimators in
samples containing fewer than $500$ observations. In samples of at least $500$
observations, all but the testing procedure based on $\hat{\Psi}_\text{CFTMLE}$
correctly reject the homogeneous treatment effect hypothesis in virtually all
replicates. As the measurement error probability increases, however, the
empirical power of the causal forest omnibus test steadily decreases ---
especially in smaller sample sizes --- whereas the empirical power of the tests
based on the absolute differential variance estimators is minimally affected.
Of note, the testing procedures based on $\hat{\Psi}_\text{TMLE}$,
$\hat{\Psi}_\text{OS}$, and $\hat{\Psi}_\text{CFOS}$ have near identical
empirical power in all scenarios and sample sizes, whereas the test based on
$\hat{\Psi}_\text{CFTMLE}$ has noticeably worse performance in all data sets
with fewer than $1000$ observations. As expected, when the true effect modifier
is completely mismeasured, only the tests based on the differential variance
estimands can reliably detect heterogeneous treatment effects. Indeed,
$\hat{\Psi}_\text{TMLE}$, $\hat{\Psi}_\text{OS}$, and $\hat{\Psi}_\text{CFOS}$
have an empirical power above $80\%$ even when $n = 250$ in this setting.
Conversely, the omnibus test based on the causal forest has negligible power
regardless of sample size.

These results demonstrate that differential variance inference procedures can
reliably detect heterogeneous treatment effects in settings where CATE-based
procedures may lack power due to, for example, measurement error. Additionally,
this experiment provides empirical evidence that the proposed approach can uncover
violations of the second-moment homogeneous treatment effect hypothesis when all
effect modifiers are omitted from the data.

\section{TTM and TTM2 Trial Analyses}\label{sec:application}

We now turn our attention to the motivating application, having demonstrated the
differential variance estimators' desirable behaviour in finite-sample
simulations. Recall that TTM and TTM2 trials investigated the optimal
temperature for targeted temperature management of out-of-hospital cardiac
arrest patients. The TTM and TTM2 trials respectively randomized $950$ and
$1900$ unconscious adult patients admitted to participating hospitals following
an out-of-hospital cardiac arrest to one of two intervention arms. In the TTM
trial, patients were randomized with equal probability to targeted temperature
management at either \SI{33}{\celsius} or \SI{36}{\celsius}. In the TTM2 trial,
patients were randomized with equal probability to either targeted temperature
management at \SI{33}{\celsius} or targeted normothermia, which aims to maintain
patients' body temperature below \SI{37.8}{\celsius}. While we remind 
that these trials produced null results with respect to patients' marginal six-month survival,
\citet{markPersonalizingTemperatureTargets2025} recently posited that patients with
severe neurological injury or prolonged ischemia duration may derive increased
benefit from lower target temperatures. A forthcoming analysis targeting the CATE
and TEM-VIPs failed to provide evidence of effect heterogeneity, however, possibly
due to missing or mismeasured effect modifiers \citep{lileikyte2026}.

We applied our absolute differential variance procedure as a sensitivity
analysis to investigate departures from the second-moment homogeneous treatment
effect hypothesis. The previously outlined conditions required for valid causal
inference of this estimand are met by virtue of the data being generated in an RCT. While
this implies that unadjusted estimators of the absolute differential variance
could be employed, we again emphasize that the adjusted estimators developed
here are generally more efficient \citep{zaki2026}. Second-moment homogeneous
treatment effect hypothesis tests based on adjusted estimators are
therefore expected to have greater power than one based on an unadjusted estimator,
an important consideration given that the TTM and TTM2 trials were not designed
to detect a heterogeneous treatment effect. Thus, guided by the numerical
experiment results presented in the third simulation study of
Section~\ref{sec:simulations} and using an identical nuisance parameter estimation
strategy, we estimated the absolute differential variances of the truncated survival
times across the trials' respective treatment groups using the $\Psi_\text{OS}$ and
$\Psi_\text{TMLE}$. These estimators adjusted for age, biological sex, time to return
of spontaneous circulation, time to advanced life saving support, presence of circulatory
shock on admission, initial cardiac rhythm, adrenaline administration, whether the
cardiac arrest was witnessed by a bystander, presence of bystander
cardiopulmonary resuscitation, and functional outcome before cardiac arrest
(defined according to Cerebral Performance Category in the TTM trial and
modified Rankin Scale score in the TTM2 trial). We additionally adjusted for patients'
Charlson Comorbidity Index in the TTM2 trial data analysis
\citep{charlsonNewMethodClassifying1987}, though this was not available in the
TTM data. We note that we treated the six-month survival times as continuous outcomes
given the virtual lack of right-censoring. The results are presented in
Table~\ref{tab:ttm2-results}.

\begin{table}
  \centering
  \caption{TTM and TTM2 Trial Absolute Differential Variance Inference Results}
  \label{tab:ttm2-results}
  \begin{tabular}{|l|l|c|c|c| c|}
    \hline
    Trial & Estimator & Estimate (days) & Standard Error & $95\%$ Confidence Interval & P-Value \\
    \hline
    \hline
    TTM & One-Step & 3.89 & 7.08 & (-9.98, 17.77) & 0.58 \\
    \hline
    TTM & TMLE & 3.61 & 7.08 & (-10.27, 17.49) & 0.61 \\
    \hline
    TTM2 & One-Step & 0.87 & 0.33 & (0.22, 1.51) & <0.01 \\
    \hline
    TTM2 & TMLE & 0.88 & 0.33 & (0.23, 1.53) & <0.01 \\
    \hline
  \end{tabular}
\end{table}

While the second-moment homogeneous treatment effect hypothesis cannot be
rejected based on the TTM trial data --- perhaps owing to its small sample size
relative to TTM2 --- it is rejected in the TTM2 trial. Inference about the
absolute differential variance using both $\Psi_\text{OS}$ and
$\Psi_\text{TMLE}$ provides evidence of a heterogeneous response to the
interventions considered in the TTM2 trial (Table~\ref{tab:ttm2-results}). The
effect estimate is small, however: the estimated difference of the potential
outcomes' standard deviations is less than one day. Coupled with the TTM2
trial's original null overall findings and the analyses of
\citet{lileikyte2026} inability to refute $f^{(1)}(W) = f^{(0)}(W) + \gamma$
based on many routinely collected patient characteristics, these results
suggest that the heterogeneity in patient response hypothesized by
\citet{markPersonalizingTemperatureTargets2025} is not likely to be clinically
actionable even though there is evidence of its existence. That is, one target temperature should not be favored over another. This analysis highlights
the sensitivity of the proposed methodology for detecting heterogeneous treatment
effects, as well as the importance of generating interpretable statistical evidence
that translates straightforwardly to clinical decision-making.

\section{Discussion}\label{sec:discussion}

Motivated by the timely re-analysis of clinical trials in critical care and the
limitations of standard approaches for detecting heterogeneous treatment
effects, we proposed causally interpretable differential variance estimands and
showed that these contrasts can be used to detect violations of a relaxed but
practical homogeneous effect hypothesis without conditioning on effect
modifiers. We then established conditions under which
these causal estimands are estimable in observed-data DGPs and derived
accompanying assumption-lean estimators. We studied these estimators'
asymptotic properties --- theoretically and empirically --- showing that they
are double robust and asymptotically linear under mild conditions about the
DGP and nuisance parameter estimators. These results, in turn, facilitate
straightforward hypothesis testing about treatment effect heterogeneity in
randomized and observational studies where effect modifiers are possibly
missing or mismeasured.

We are not the first to consider outcome variances
as means to detect treatment effect heterogeneity. \citet{coxInteraction1984}
provided a broad review of statistical interactions, noting that two treatment
groups sharing identical outcome variances represent ``a special form of absence
of interaction.'' Though a causal framework and modern terminology were not employed,
\citet{coxInteraction1984} implicitly equated constant potential outcome variances
to the absence of effect modifiers --- and therefore
treatment effect homogeneity. Later, \citet{dingRandomizationInferenceTreatment2016}
briefly discussed the ratio of potential outcome variances --- here, the relative 
differential variance --- to test the shifted sharp homogeneous treatment effect
hypothesis in the context of design-based inference strategies for uncovering
heterogeneity. They nevertheless considered the traditional F-test for randomized
studies and derived an asymptotically distribution-free alternative that relaxes
this test's normality assumption. More recently, \citet{millsDetectingHeterogeneityIntervention2021} benchmarked several parametric procedures for comparing RCT arms' variances. As with \citet{coxInteraction1984}, however, these comparisons were not performed within a causal framework.

Our work therefore builds on these prior contributions in several important ways.
First, we proposed a homogeneous treatment effect hypothesis that is better suited
for (nonparametric) model-based causal inference. Whereas the shifted sharp
homogeneous treatment effect hypothesis is untenable in many settings,
second-moment homogeneity captures a meaningful and practical notion of a
homogeneous treatment effect. Further, we establish how this homogeneity definition supports sensitivity analyses of null conditional heterogeneous treatment effect investigations. Second, we proposed interpretable differential
variance estimands that are straightforward for non-statisticians to incorporate
into decision-making tasks. They additionally generalize previously proposed
estimands of the potential outcome variances. Finally, we developed several
causal machine learning estimators that permit adjustment for
confounding variables, in turn allowing second-moment homogeneous treatment effect
tests to be performed in observational studies. These estimators additionally allow
adjustment for prognostic factors, thereby leading to more efficient detection of
heterogeneous treatment effects in RCTs \citep{zaki2026}.

The proposed methodology is not without its limitations. In particular, differential variance estimands do not supply information about the source
or directionality of a heterogeneous effect when present --- something
provided by CATE or TEM-VIP when effect modifiers are included in these
parameters' conditioning sets. We therefore view our approach as complementary
to standard procedures for uncovering heterogeneity, as demonstrated
in the motivating sensitivity analysis of Section~\ref{sec:application}.

Our contributions give rise to several exciting avenues of research. First, our 
proposed methodology can be adapted to DGPs with diverse treatment assignment
mechanisms and outcome types, permitting formal testing of homogeneous
treatment effect hypotheses in myriad settings where treatment effect modifiers
are presumed missing from available data. Second, our potential outcome variance
estimators could extend the work of 
\citet{volkmannRelationshipTreatmentEffect2020} and \citet{galanterCanPotentialBenefit2024},
which provide bounds for the CATE in RCTs based solely on the arms'
outcome variances, to observational studies. The resulting bounds could therefore
help determine whether a resource-intensive conditional heterogeneous treatment 
effect analysis of a non-randomized study is worthwhile. The use of our
estimators when deriving these bounds with RCT data would be equally
advantageous owing to their generally increased precision over unadjusted
estimators. Third, future work might also combine our inference procedures
with the nonparametric CATE variance estimators of\citet{levyFundamentalMeasureTreatment2021} to directly target contrasts
of residual potential outcome variances, like
$\mathbb{E}_{P_{X,0}}[\mathbb{V}_{P_{X,0}}[\epsilon^{(1)}|W]] - \mathbb{E}_{P_{X,0}}[\mathbb{V}_{P_{X,0}}[\epsilon^{(0)}|W]]$.
While causal machine learning estimators of these estimands would not
possess the robustness properties of the estimators derived here, they could
nevertheless be used to quantify heterogeneity unexplained by available, known effect modifiers. In assessing the adequacy of available covariates for explaining treatment effect variation, inference about residual potential outcome variances could inform the interpretation and application of CATE estimates in clinical practice.

\section*{Software and Code}

The differential variance methodology proposed in this manuscript is implemented
in the open-source \texttt{cmldiffvar} software for the R Language for
Statistical Computing \citep{rlang}. This package is available on GitHub at
\url{https://github.com/PhilBoileau/cmldiffvar}.

The code for the simulation studies of Section~\ref{sec:simulations} is also
freely available at
\url{https://github.com/PhilBoileau/pub_differential-variance-simulations}.
These numerical experiments rely on the simulation study framework implemented
in the \texttt{simChef} R package \citep{Duncan2024}.

\section*{Acknowledgments}

The authors thank Nima S. Hejazi for his insightful comments and suggestions on
an early version of the manuscript. PAB is supported by a Natural Sciences and 
Engineering Research Council of Canada Discovery Grant. GL reports financial
support by the Royal Physiographic Society of Lund and the Swedish Society for
Medical Research. MES is a Tier 2 Canada Research Chair in Causal Inference and
Machine Learning in Health Science and is supported by a Natural Sciences and
Engineering Research Council of Canada Discovery Grant.

\nopagebreak
\putbib

\end{bibunit}

\newpage

\appendix

\begin{bibunit}

\section{Proofs}

\subsection{Theorem~\ref{thm:constant-ite-violations-abs-diff-var}}

\begin{proof}
  We present a proof by contradiction. Let $\Psi_F(P_{X,0}) \neq 0$ and assume
  $f^{(0)}(W) = f^{(1)}(W) - \gamma$ and
  $\mathbb{E}_{P_{X,0}}[\mathbb{V}_{P_{X,0}}[\epsilon^{(0)}|W]] = \mathbb{E}_{P_{X,0}}[\mathbb{V}_{P_{X,0}}[\epsilon^{(1)}|W]]$. Then

    \begin{equation*}
        \begin{split}
          \Psi_F(P_{X,0})
          & = \sqrt{\mathbb{V}_{P_{X,0}}[Y^{(1)}]} - \sqrt{\mathbb{V}_{P_{X,0}}[Y^{(0)}]} \\
          & = \sqrt{\mathbb{V}_{P_{X,0}}\left[f^{(1)}(W)\right] + \mathbb{E}_{P_{X,0}}\left[\mathbb{V}_{P_{X,0}}\left[\epsilon^{(1)}|W\right]\right]} -\\
          & \qquad \qquad \sqrt{\mathbb{V}_{P_{X,0}}\left[f^{(0)}(W)\right] + \mathbb{E}_{P_{X,0}}\left[\mathbb{V}_{P_{X,0}}\left[\epsilon^{(0)}|W\right]\right]} \\
          & = \sqrt{\mathbb{V}_{P_{X,0}}\left[f^{(0)}(W) + \gamma \right] + \mathbb{E}_{P_{X,0}}\left[\mathbb{V}_{P_{X,0}}\left[\epsilon^{(0)}|W\right]\right]} -\\
          & \qquad \qquad \sqrt{\mathbb{V}_{P_{X,0}}\left[f^{(0)}(W)\right] + \mathbb{E}_{P_{X,0}}\left[\mathbb{V}_{P_{X,0}}\left[\epsilon^{(0)}|W\right]\right]} \\
          & = \sqrt{\mathbb{V}_{P_{X,0}}\left[f^{(0)}(W)\right] + \mathbb{E}_{P_{X,0}}\left[\mathbb{V}_{P_{X,0}}\left[\epsilon^{(0)}|W\right]\right]} -\\
          & \qquad \qquad \sqrt{\mathbb{V}_{P_{X,0}}\left[f^{(0)}(W)\right] + \mathbb{E}_{P_{X,0}}\left[\mathbb{V}_{P_{X,0}}\left[\epsilon^{(0)}|W\right]\right]} \\
          & = 0
        \end{split}
    \end{equation*}
    The first equality follows from the law of total variance. The second
    equality follows by assumption. The third equality follows from the location 
    invariance of the variance. Thus, at least one of the conditions is violated.
\end{proof}

\subsection{Theorem~\ref{thm:moment-identifiability}}
\begin{proof}
    \begin{equation*}
        \begin{split}
            \mu^d(P_{X,0}; a)
            & = \mathbb{E}_{P_{X,0}}[\mathbb{E}_{P_{X,0}}[(Y^{(a)})^d|W]] \\
            & = \mathbb{E}_{P_{X,0}}[\mathbb{E}_{P_{X,0}}[(Y^{(a)})^d|W, A=a]] \\
            & = \mathbb{E}_{P_0}[\bar{Q}_{P_0}^d(W, a)] \\
            & = \mu^d(P_{0}; a) \;
        \end{split}
    \end{equation*}
\end{proof}

\subsection{Proposition~\ref{prop:sigma-eif}}

\begin{proof}
  Given that the EIF is a functional derivative and that
  $\sigma^2(P; a) = \mu^2(P; a) - \mu(P; a)^2$, we apply the functional delta
  method to obtain $D^{\sigma^2(a)}(P; O)$:
  \begin{equation*}
    D^{\sigma^2(a)}(P; O) = D^{\mu^2(a)}(P; O) - 2\mu(P; a)D^{\mu(a)}(P; O) \;,
  \end{equation*}
  where $D^{\mu^2(a)}(P;O)$ and $D^{\mu(a)}(P;O)$ are the EIFs of $\mu^2(P;a)$
  and $\mu(P;a)$, respectively. Using the point-mass contamination approach with
  $P_{t} = t\tilde{P} + (1-t)P$ for $t \in [0, 1]$ where $\tilde{P}$ is a
  degenerate distribution with point mass at $\tilde{o}$, the EIF of
  $\mu^d(a)$ is derived as follows:
  \begin{equation*}
    \begin{split}
      \frac{d\mu^d(P_{t};a)}{dt} \bigg|_{t=0}
      & = \frac{d}{dt}\left\{\int y^d f_t(y|a, w)f_t(w)dydw \right\} \bigg|_{t=0} \\
      & = \frac{d}{dt}\left\{ \int y^d \frac{f_t(y, a, w)f_t(w)}{f_t(a,w)}dydw \right\} \bigg|_{t=0} \\
      & = \int \frac{d}{dt}\left\{y^d \frac{f_t(y, a, w)f_t(w)}{f_t(a,w)}\right\}\Bigg|_{t=0}dydw \\
      & = \int y^d \Bigg[\frac{f(w)}{f(a,w)}\frac{d}{dt}\left\{f_t(y, a, w)\right\}\Bigg|_{t=0}
        - \frac{f(y, a, w)f(w)}{f(a,w)^2}\frac{d}{dt}\left\{f(a,w)\right\}\Bigg|_{t=0} +\\
      & \qquad\qquad \frac{f(y, a, w)}{f(a,w)}\frac{d}{dt}\left\{f_t(w)\right\}\Bigg|_{t=0}
        \Bigg] dydw \\
      & = \int y^d \frac{f(y, a, w)f(w)}{f(a,w)}
        \Bigg[
        \frac{\delta_{\tilde{y}, \tilde{a}, \tilde{w}}(y, a, w)}{f(y, a, w)}
        - \frac{\delta_{\tilde{a}, \tilde{w}}(a, w)}{f(a, w)} +
        \frac{\delta_{\tilde{w}}(w)}{f(w)} - 1 \Bigg] dydw \\
      & = \int \frac{\tilde{y}^d\delta_{\tilde{y}, \tilde{a}, \tilde{w}}(y, a, w)}{f(a|\tilde{w})}
        - \frac{y^df(y|a, \tilde{w})\delta_{\tilde{a}, \tilde{w}}(a, w)}{f(a|\tilde{w})} + y^df(y|a, \tilde{w})\delta_{\tilde{w}}(w) -\\
      & \qquad \qquad y^d f(y|a, w)f(w) dydw \\
      & = \frac{\tilde{y}^d \delta_{\tilde{a}}(a)}{I(A = 1)g_P(\tilde{w}) + I(A=0)(1-g_P(\tilde{w}))} - \frac{\delta_{\tilde{a}}(a) \bar{Q}^d_P(\tilde{w}, a)}{I(A = 1)g_P(\tilde{w}) + I(A=0)(1-g_P(\tilde{w}))} +\\
      & \qquad\qquad \bar{Q}^d_P(\tilde{w}, a) - \mu^d(P;a) \; ,
        \end{split}
    \end{equation*}
    where $\delta_{\tilde{o}}$ is the Dirac delta function evaluated at $\tilde{o}$.

    Then
    \begin{equation*}
        D^{\mu^d(a)}(P;O) = \frac{I(A = a)\left(Y^d - \bar{Q}^d_P(W, a)\right)}{I(A = 1)g_P(W) + I(A=0)(1-g_P(W))} + \bar{Q}^d_P(W, a) - \mu^d(P;a) \;,
    \end{equation*}
    and the result follows.
\end{proof}

\subsection{Theorem~\ref{thm:double-robustness}}

\begin{proof}
  The TMLE, one-step estimators, and their cross-fitted variants are
  asymptotically equivalent \citep{vanderlaanTargetedLearningCausal2011}.
  Consistency of these estimators can therefore be studied by evaluating the
  finite bias of the one-step estimator. Let $c < g_n(W) < 1-c$ a.s. for some
  $c \in (0, 1)$. Then,
  \begin{equation*}
    \begin{split}
      \text{Bias}(\sigma^2_\text{OS}(\hat{P}_n; 1))
      & = \mathbb{E}_{P_0}\left[\sigma^2_\text{OS}(\hat{P}_n; 1) - \sigma^2(P_0; 1)\right] \\
      & = \mathbb{E}_{P_0}\bigg[\frac{A}{g_n(W)}\left(Y^2 - 2Y\mu_n(1) + 2\bar{Q}_n(W, 1)\mu_n(1) - \bar{Q}_n^2(W, 1)\right) +\\
      & \qquad \qquad  \bar{Q}_n^2(W,1) - 2\bar{Q}_n(W,1)\mu_n(1) + \mu_n(1)^2 - \sigma^2(P_0; 1)\bigg] \\
      & = \mathbb{E}_{P_0}\bigg[\frac{A}{g_n(W)}\left(Y^2 - 2Y\mu_n(1) + 2\bar{Q}_n(W, 1)\mu_n(1) - \bar{Q}_n^2(W, 1)\right) +\\
      & \qquad \qquad  \bar{Q}_n^2(W,1) - 2\bar{Q}_n(W,1)\mu_n(1) + \mu_n(1)^2 - \bar{Q}^2_{P_0}(W, 1) + \mu(P_0; 1)^2\bigg]  \\
      & = \mathbb{E}_{P_0}\bigg[\frac{A}{g_n(W)}\left(Y^2 - 2Y\mu_n(1) + 2\bar{Q}_n(W, 1)\mu_n(1) - \bar{Q}_n^2(W, 1) \right) - \\
      & \qquad \qquad \left(\bar{Q}^2_{P_0}(W, 1) - 2\bar{Q}_{P_0}(W, 1)\mu_n(1) + 2\bar{Q}_n(W,1)\mu_n(1) - \bar{Q}_n^2(W,1)\right)\bigg] +\\
      & \qquad\qquad \left(\mu_n(1) - \mu(P_0; 1)\right)^2 \\
      & = \mathbb{E}_{P_0}\bigg[\frac{g_{P_0}(W)}{g_n(W)}\left(\bar{Q}^2_{P_0}(W,1) - 2\bar{Q}_{P_0}(W,1)\mu_n(1) + 2\bar{Q}_n(W, 1)\mu_n(1) - \bar{Q}_n^2(W, 1) \right) - \\
      & \qquad \qquad \left(\bar{Q}^2_{P_0}(W, 1) - 2\bar{Q}_{P_0}(W, 1)\mu_n(1) + 2\bar{Q}_n(W,1)\mu_n(1) - \bar{Q}_n^2(W,1)\right)\bigg] +\\
      & \qquad\qquad  \left(\mu_n(1) - \mu(P_0; 1)\right)^2 \\
      & = \mathbb{E}_{P_0}\bigg[\frac{g_{P_0}(W)-g_n(W)}{g_n(W)}\left(\bar{Q}^2_{P_0}(W,1)- \bar{Q}_n^2(W, 1)\right) +\\
      & \qquad \qquad 2\mu_n(1)\frac{g_{P_0}(W)-g_n(W)}{g_n(W)}\left(\bar{Q}_n(W,1) - \bar{Q}_{P_0}(W,1)\right)\bigg] +\\
      & \qquad \qquad \left(\mu_n(1) - \mu(P_0; 1)\right)^2\;.\\
    \end{split}
  \end{equation*}

  An analogous result is obtained when $a = 0$. It then follows that
  $\sigma^2_\text{OS}(\hat{P}_n; a)$ is consistent under the specified
  conditions.
\end{proof}

\subsection{Theorem~\ref{thm:asymptotic-linearity}}

\begin{proof}
  The asymptotic linearity of the nonparametric estimators is established
  through the standard study of the empirical process term and the second order remainder
  term of the von Mises expansion
  \citep{vanderlaanTargetedLearningCausal2011, hinesDemystifyingStatisticalLearning2022, kennedyOptimalDoublyRobust2023}. By Lemma~2 of
  \citet{kennedySharpInstrumentsClassifying2020}, we find that the empirical
  process term is $o_P(n^{-1/2})$ under the first and second assumptions. Next, letting
  $c < g_n(W) < 1-c$ a.s. for some $c \in (0, 1)$, we find that:
  \begin{equation*}
    \begin{split}
      \lvert R(P_0, \hat{P}_{n}) \rvert
      & = \left\lvert \sigma^2(\hat{P}_{n}; 1) - \sigma^2(P_0; 1) + \mathbb{E}_{P_0}[D^{\sigma^2(1)}(\hat{P}_{n}; O)] \right\rvert \\
      & = \bigg\lvert\mathbb{E}_{P_0}\bigg[\frac{g_{P_0}(W)-g_n(W)}{g_n(W)}\left(\bar{Q}^2_{P_0}(W,1)- \bar{Q}_{n}^2(W, 1)\right) +\\
      & \qquad \qquad 2\mu_n(1)\frac{g_{P_0}(W)-g_n(W)}{g_n(W)}\left(\bar{Q}_{n}(W,1) - \bar{Q}_{P_0}(W,1)\right)\bigg]\bigg\rvert + \left(\mu_{n}(1) - \mu(P_0; 1)\right)^2 \\
      & \leq \frac{1}{c}\mathbb{E}_{P_0}\big[\left(g_{P_0}(W)-g_{n}(W)\right)^2\big]^{1/2}\mathbb{E}_{P_0}\big[\left(\bar{Q}^2_{P_0}(W,1) - \bar{Q}_{n}^2(W, 1)\right)^2\big]^{1/2} + \\
      & \qquad \qquad \frac{2}{c}\lvert\mu_{n}(1)\rvert\mathbb{E}_{P_0}\big[\left(g_{P_0}(W) - g_{n}(W)\right)^2\big]^{1/2}\mathbb{E}_{P_0}\big[\left(\bar{Q}_{n}(W,1) - \bar{Q}_{P_0}(W,1)\right)^2\big]^{1/2} + \\
      & \qquad \qquad \qquad \left(\mu_{n}(1) - \mu(P_0; 1)\right)^2 \;. \\
    \end{split}
  \end{equation*}
  The second equality follows from the proof of
  Theorem~\ref{thm:double-robustness} and the inequality from the Cauchy-Schwarz
  inequality. It follows that $\sqrt{n}R(P_0, \hat{P}_{n}) = o_P(1)$. Analogous
  results are obtained when $a = 0$ and for the cross-fitted estimators.
\end{proof}

\nopagebreak
\putbib

\end{bibunit}

\end{document}